\documentclass[journal=jacsat,manuscript=article]{achemso}

\usepackage[version=3]{mhchem} 
\usepackage{cleveref}

\author{Lucia Baldauf}
\affiliation{Department of Bionanoscience, Kavli Institute of Nanoscience Delft, Delft University of Technology, 2629 HZ Delft, The Netherlands}
\altaffiliation{Equal contributions}
\author{Lennard van Buren}
\affiliation{Department of Bionanoscience, Kavli Institute of Nanoscience Delft, Delft University of Technology, 2629 HZ Delft, The Netherlands}
\altaffiliation{Equal contributions}
\author{Federico Fanalista}
\affiliation{Department of Bionanoscience, Kavli Institute of Nanoscience Delft, Delft University of Technology, 2629 HZ Delft, The Netherlands}
\author{Gijsje Hendrika Koenderink}
\affiliation{Department of Bionanoscience, Kavli Institute of Nanoscience Delft, Delft University of Technology, 2629 HZ Delft, The Netherlands}
\altaffiliation{Corresponding author}
\email{g.h.koenderink@tudelft.nl}

\title[]
  {Actomyosin-driven division of a synthetic cell}

\abbreviations{}

\begin{document}


\begin{abstract}
    One of the major challenges of bottom-up synthetic biology is rebuilding a minimal division machinery. The animal cell division apparatus is mechanically the simplest, in which an actin-based ring constricts the membrane, as compared to microbes and plant cells where a cell wall is involved. Furthermore, reconstitution of the actin division machinery helps to understand the physical and molecular mechanisms of cytokinesis in animal cells and thus our own cells. In this review, we describe the state-of-the-art research on reconstitution of minimal actin-mediated cytokinetic machineries. Based on the conceptual requirements that we obtained from the physics of the shape changes involved in cell division, we propose two major routes for building a minimal actin apparatus capable of division. Importantly, we acknowledge both the passive and active roles that the confining lipid membrane can play in synthetic cytokinesis. We conclude this chapter by identifying the most pressing challenges for future reconstitution work, thereby laying out a roadmap for building a synthetic cell equipped with a minimal actin division machinery.
\end{abstract}

\section{Introduction}
Bottom-up synthetic biology is an emerging field at the interface of cell biology, (bio)chemistry and (bio)physics. Several national and international initiatives have been founded recently, which are aimed at reconstituting synthetic cells that can autonomously grow and divide \cite{Staufer2021b, Frischmon2021}. As a chassis, usually giant unilamellar vesicles (GUVs) are used, which are cell-sized (5-50 \textmu m) containers enveloped in a lipid bilayer \cite{Mulla2018, Spoelstra2018, Gaut2021, Litschel2021b}. One of the key functions that a synthetic cell must be able to perform in order to be considered life-like is cytokinesis \cite{Olivi2021}, a process in which a cell physically splits into two daughter cells. To reconstitute cytokinesis, various strategies are being pursued, inspired by biological strategies employed by prokaryotic, archaeal or eukaryotic cells \cite{Kretschmer2019, Olivi2021}. These biological systems have in common that cell division is accomplished by a cytoskeletal protein machinery, often ring-shaped, that assembles at the cell equator. In microbial cells (bacteria and yeast), this protein machinery collaborates with a complex cell wall synthesis machinery \cite{Mahone2020, Wang2020}. By contrast, animal cells lack a cell wall and cytokinesis is entirely driven by the actin cytoskeleton. Actin-based cell division could thus be an ideal basis for engineering synthetic cell division. 

Bottom-up reconstitution of actin-based cell division is interesting not only from an engineering perspective, but also as a means to understand how cytokinesis works at the molecular level in animal cells. Although cytokinesis is a well-studied cellular process, surprisingly many fundamental questions about its working principles remain unanswered \cite{Pollard2017}: how are mechanical forces generated and sustained? How much molecular complexity is needed to ensure that the actin cortex retains its structural integrity  during cytokinesis? What are the requirements for cortex-membrane interactions to promote furrow ingression? These questions are difficult to address in cell-based studies because of the enormous molecular complexity of cells combined with substantial variation between cytokinetic mechanisms employed by different species and different cell types \cite{Cortes2018, Leite2019, Wang2020}. 

In this review, we propose a roadmap towards the bottom-up reconstitution of actin-driven cytokinesis in minimal cells. For brevity, we consider only the process of furrow ingression, neglecting other aspects such as membrane abscission and chromosome and cytoplasmic segregation, which are reviewed elsewhere \cite{Jongsma2015, Addi2018, Horvath2020, Anjur-Dietrich2021}. Based on theoretical models of cytokinesis in animal cells, we first identify four central biophysical requirements for actin-driven furrow ingression. Next we review experimental insights obtained from recent efforts to reconstitute minimal actin systems. We also emphasize the importance of controlling the surface area of the synthetic plasma membrane to enable cell division. Finally we propose a roadmap towards building a molecular machinery that can successfully deform a minimal cell-like container. 

\section{Biophysical requirements for making a cell divide}
\label{section:rev-requirements}

\begin{figure}[h!]
	\centering
	\includegraphics[width=0.8\textwidth]{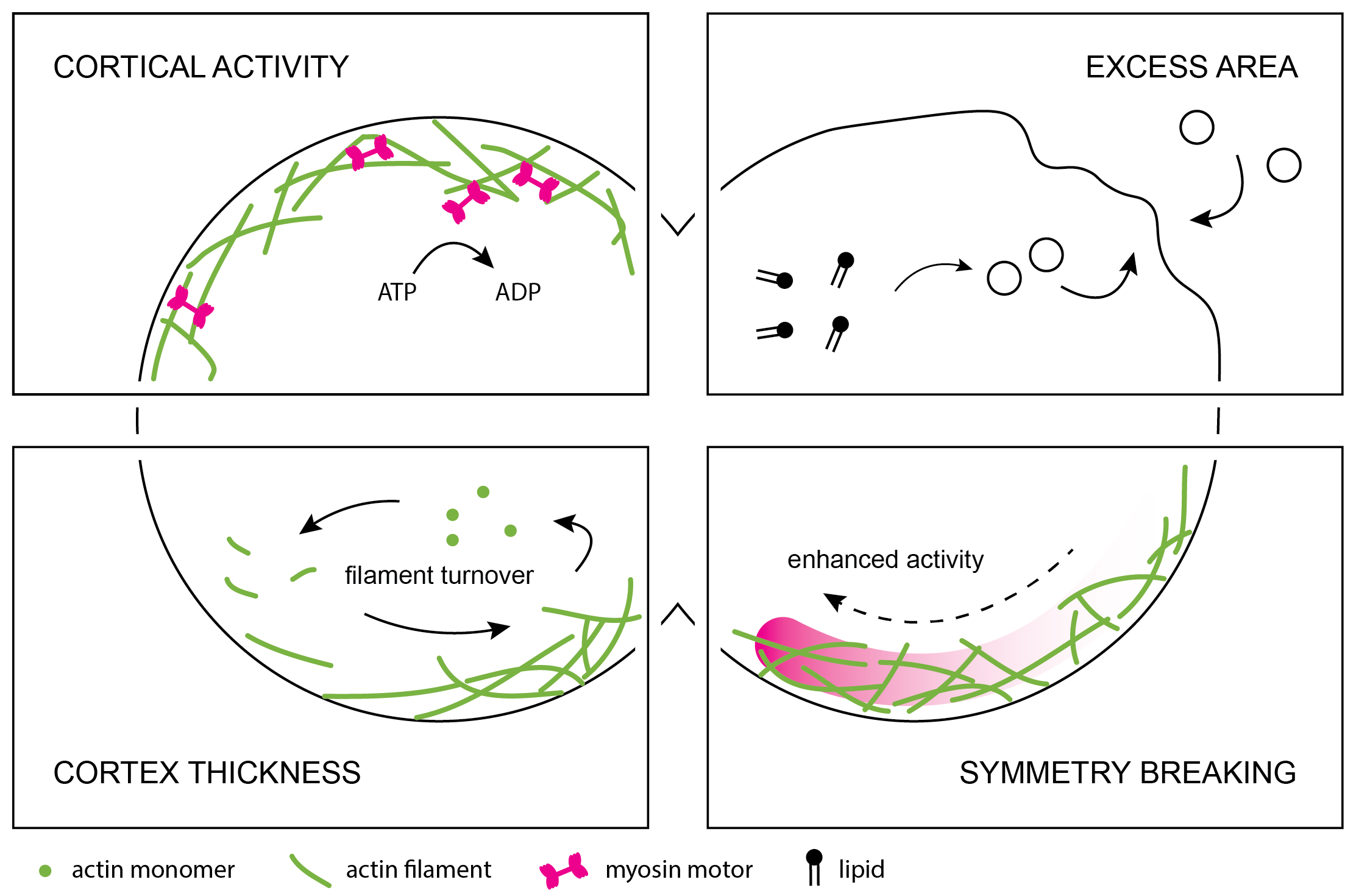}
	\caption{\textbf{Four key biophysical requirements for reconstituting synthetic cell division.} For cell deformation to occur, cortical activity driven by ATP hydrolysis is required (top left), which can for example be generated by myosin activity. Regulation of cortex thickness (bottom left) is essential for control of cortical activity and is determined by the rate of actin filament turnover versus cortical flows. For cortical activity to lead to cell deformation, the symmetry of the system needs to be broken (bottom right). Finally, to accommodate the drastic change in surface-to-volume ratio during cell division, excess membrane area needs to be generated prior to or during cytokinesis (top right).}
	\label{fig:review-conceptualisation}
\end{figure}

Cytokinesis in animal cells is a complicated process that involves many different molecular components (lipids and proteins) whose interactions and localization are tightly regulated. At a coarse-grained level, however, it is possible to formulate general biophysical requirements for cell division based on a consideration of the mechanical forces at play. Pioneering experimental work from the 1950s onward has demonstrated that cytokinesis is accompanied by membrane furrowing \cite{Roberts1961}, cortical stiffening \cite{Mitchison1954, Wolpert1966} and the appearance of ordered filamentous structures in the cytokinetic ring \cite{Arnold1969, Schroeder1968}. These observations have served as input for coarse-grained theoretical and computational models that describe cytokinesis as the shape evolution of a thin, viscoelastic and active shell around a (nearly) constant volume of cytoplasm. From the models, we can infer several key requirements that a cell, living or synthetic, must fulfil in order to successfully divide (\cref{fig:review-conceptualisation}):

\noindent \textbf{1. Cortical activity.} The actin cortex driving cytokinesis in animal cells must be active. This means that it should include elements that hydrolyse adenosine triphosphate (ATP), an energy-carrying nucleotide, to generate contractile forces that produce cellular shape changes. The viscoelastic and active nature of the cortex can be described using the framework of active gel theory as proposed by Kruse et al. \cite{Kruse2005}. This formalism is usually applied in the viscous limit \cite{Salbreux2009, Zumdieck2005, Turlier2014, Reymann2016}, as cytokinesis is slow (minutes) compared to the fluidization time scale (10 s) of the actin cortex \cite{Salbreux2009}. The molecular origins of active force production are complex and depend on molecular detail, as discussed below.

\noindent \textbf{2. Cortical thickness.} Cortex activity, at least when mediated by myosin motors, is roughly proportional to cortical thickness \cite{Zumdieck2005, Salbreux2009, Turlier2014}. To maintain cortical activity, the cortex must consequently be of a controlled thickness. Cortical thickness is regulated by a balance of actin polymerization and depolymerization, or turnover, and cortical flows: cortical flows accumulate material in the cytokinetic furrow, whereas turnover redistributes actin throughout the cell. This suggests two requirements for synthetic cell division. Firstly, components of the cortex must be laterally mobile to be effectively redistributed by cortical flows \cite{Salbreux2009, Turlier2014, White1983}. Secondly, actin turnover rates must be low enough to allow local actin accumulation and therefore increased contractility in the furrow region. If actin is removed too rapidly, furrow constriction slows down significantly and may be halted altogether \cite{Turlier2014}. On the other hand, complete lack of filament turnover in a 2D actomyosin cortex is theoretically predicted to lead to irreversible clustering of actin, inhibiting effective stress generation \cite{Hiraiwa2016}. This prediction has yet to be reconciled with experimental evidence from yeast cells, which suggests that the persistent presence of filamentous actin, rather than turnover, is key for successful contraction of the cytokinetic ring \cite{Chew2017}.

\noindent \textbf{3. Cortical symmetry breaking.} From the 1930s onwards, various models have been proposed to explain the mechanical basis of cytokinesis. The early models range from active expansion of the cell poles \cite{Swann1958}, through active pushing by the mitotic spindle\cite{Dan1948}, to spindle-mediated relaxation of the cell poles \cite{Wolpert1960, White1983} and finally active constriction of the cytokinetic furrow \cite{Marsland1950, Yoneda1972, Turlier2014, Schroeder1968}. While details vary widely between these models, they share a key characteristic: they all posit that there must be a difference in activity between the polar and equatorial regions to drive furrow ingression. After decades of research it is now widely accepted (reviewed e.g. in \cite{Green2012}) that the main driving factor of animal cell cytokinesis is actin-based constriction at the cleavage furrow. However, \textit{in vitro} reconstitution may be the ideal tool to understand actin's role in molecular detail, and to assess to which extent other mechanisms \cite{Gudejko2012, Wang2001} also contribute.

\noindent \textbf{4. Cell surface area and volume regulation.} Consistent with observations in cells, models have generally assumed that the cytoplasm is very weakly, if at all, compressible \cite{White1983, Turlier2014}. The apparent cell surface-to-volume ratio, however, changes dramatically during cytokinesis \cite{Frey2021}. It follows that the cell's (visible) surface area must be changing. In theoretical works this change in surface area is generally assumed to be energetically ‘free’, as living cells can regulate the available membrane area through a variety of processes like blebbing \cite{Sedzinski2011}, or caveolae disassembly and membrane trafficking \cite{Sinha2011, Albertson2005}. This supply of membrane on demand is probably one of the most challenging aspects to recapitulate in a reconstituted system.

\section{Roadmap towards actin-driven synthetic cell division}
\label{section:rev-roadmap}
\begin{figure}[h!]
	\centering
	\includegraphics[width=0.8\textwidth]{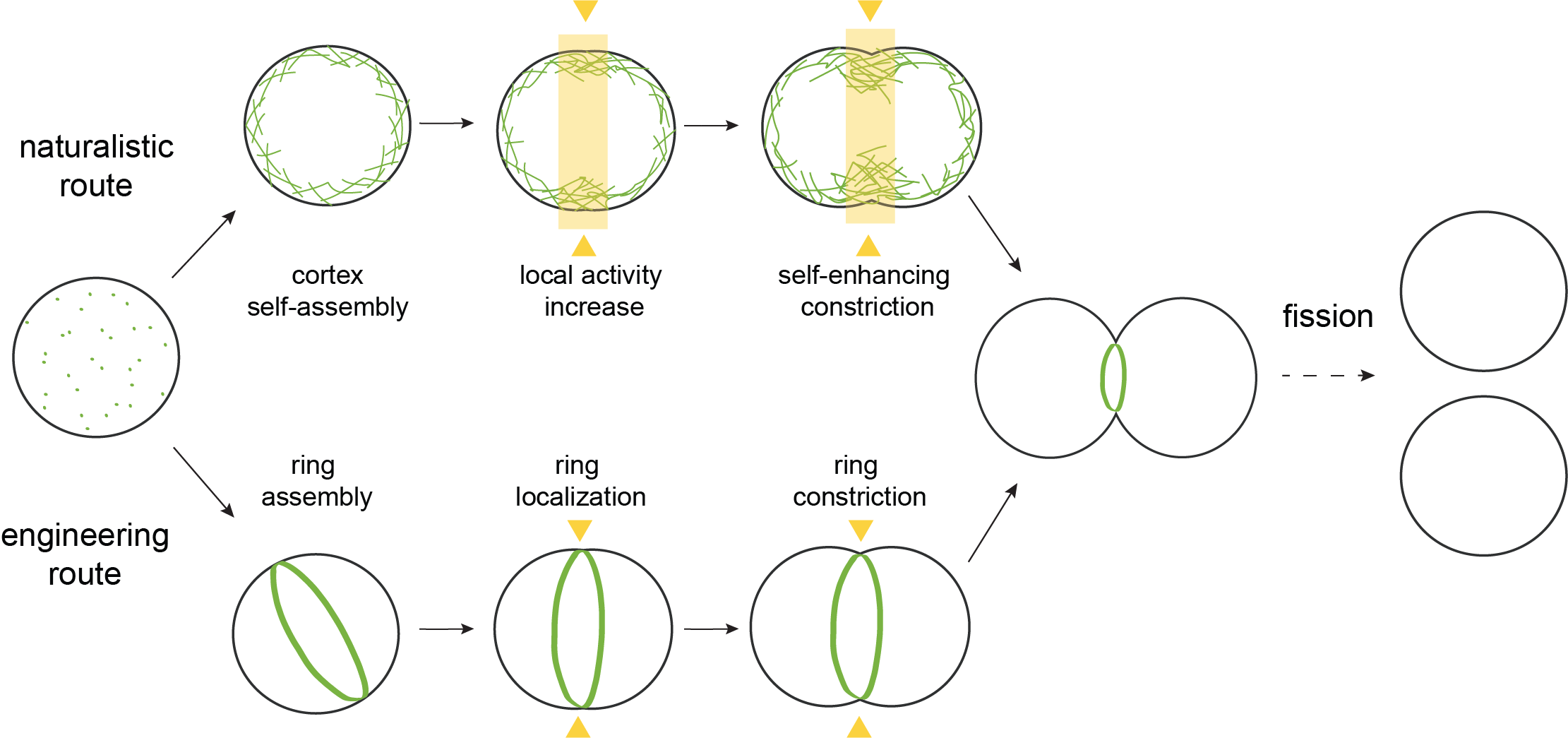}
	\caption{\textbf{Routes to actin-based synthetic cell division.} There are two main routes to achieve actin-driven division of a synthetic cell: by symmetry breaking of a reconstituted actin cortex, triggered by external or biochemical cues, which leads to self-enhanced furrow constriction (the `naturalistic' route, top), or by construction of a contractile ring at the cell equator (the `engineering' route, bottom). Yellow arrowheads indicate where contractile activity is concentrated. The final fission step is outside the scope of this review.}
	\label{fig:review-routes}
\end{figure}

Cytokinesis of animal cells is a highly complex and tightly regulated process. Yet, fairly minimal computational models are able to recapitulate cytokinesis, suggesting that the underlying mechanisms may be recreated with simplified molecular mechanisms. Here, we propose a roadmap towards reconstituting actin-driven cell division by considering lessons from recent cell and \textit{in vitro} (i.e. cell-free reconstitution) studies. Basically, there are two routes for reconstitution of actin-driven cytokinesis (see \cref{fig:review-routes}). First, cell division can be recreated via reconstitution of an actin cortex that, upon symmetry breaking, is more contractile at the cell equator as compared to the poles. This route is most close to cytokinesis in mammalian cells, and we therefore name it the naturalistic route. The second route is by construction of a cytokinetic ring that anchors and contracts at the cell equator, coined the engineering route. We will first discuss the design of an actin-based machinery fit for driving cytokinesis in both scenarios, and in the next section consider the design of the lipid membrane envelope.

\subsection{Naturalistic route: building a self-assembling cytokinetic ring}
\label{section:rev-nat}

During interphase, mammalian cells have a continuous actin cortex that lines the plasma membrane \cite{Chugh2018a}. When cells enter mitosis, the cortex is remodelled and self-assembles into a contractile ring at the cell equator. Symmetry breaking and midplane localization of the cytokinetic furrow is initiated by biochemical signalling, which includes Rho-dependent myosin phosphorylation in the furrow region \cite{Turlier2014, Asano2009}. The locally enhanced activation of myosin is thought to lead to cortical flows from the poles to the equator \cite{White1983,Bray1988}, which further accumulate and organize contractile elements in the furrow \cite{Najafabadi2022} that drive furrow ingression \cite{Turlier2014}. Such a complex self-assembling system has not been built to date, but steps have been taken along the road (\cref{fig:review-natural}). 

\subsubsection{Reconstitution of active actin networks}
Both cell-free experiments and theoretical models of cortex-like disordered actin networks have been used to elucidate why disordered actomyosin networks are contractile in the first place. The detailed mechanisms are reviewed elsewhere \cite{Koenderink2018, MendesPinto2013a, Murrell2015}, but they broadly comprise two scenarios. Actin filaments are semiflexible polymers with a thermal persistence length of 10-15 \textmu m, of the same order as their contour length \cite{Kang2012}. The first contraction scenario, relevant for well-connected networks of long filaments, is that the anisotropic mechanical force-extension response of actin filaments causes them to buckle and break under motor-induced compressive stress \cite{Murrell2012, Lenz2012}. The second scenario, relevant for networks with short actin filaments, is that the structural polarity of actin filaments in combination with the tendency of myosin II motors to dwell at the filament plus end before detachment causes contraction via polarity sorting \cite{Kruse2000, Zumdieck2005, Wollrab2019}. In the actin cortex of mammalian cells there may be a combination of both mechanisms, since distinct populations of short and long filaments are present there \cite{Fritzsche2016}. 

\begin{figure}[h!]
	\centering
	\includegraphics[width=0.8\textwidth]{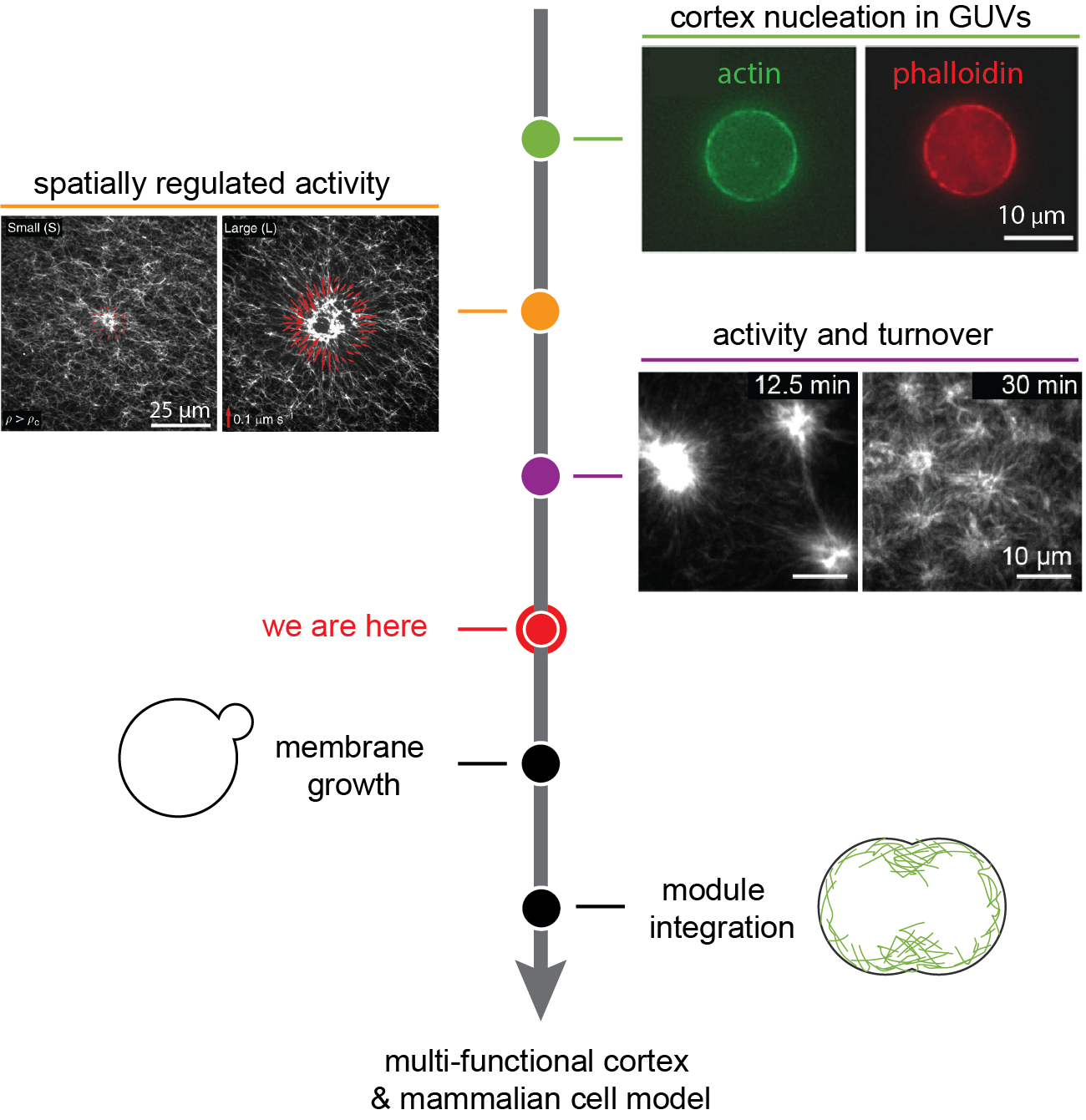}
	\caption{\textbf{Roadmap to division with an actin cortex.} Green: successful nucleation of an actin cortex inside GUVs \cite{Pontani2009a}. Filament formation of actin (green) is confirmed by co-localization of the filament-binding peptide phalloidin (red). Orange: spatiotemporal control of myosin activity by light-induced inactivation of the myosin inhibitor blebbistatin was used to generate network contraction over different length scales, from small (left) to large (right) \cite{Linsmeier2016}. Purple: combination of myosin activity with actin filament turnover generates sustained network contraction \cite{Sonal2019}. In the coming time, steps need to be taken to engineer membrane growth, and finally to integrate the different modules inside a GUV.}
	\label{fig:review-natural}
\end{figure}

Notably, the combined effect of contractile motor activity and actin turnover remains poorly explored. Theoretical models generally assume that the cytokinetic cortex does undergo actin turnover \cite{Turlier2014, Salbreux2009, Berthoumieux2014}, and have even indicated that turnover is key for sustained stress generation during furrow ingression \cite{Hiraiwa2016}. Experimentally, besides one study with a cell extract \cite{Malik-Garbi2019}, only one minimal \textit{in vitro} study has so far combined actin turnover and myosin activity \cite{Sonal2019}. This work showed that myosin activity alone can be sufficient to induce turnover in minimal actin networks (see \cref{fig:review-natural}, purple). Myosin-driven compaction and fragmentation of Arp2/3-nucleated actin led to the removal of actin from the network, and subsequent redistribution and re-incorporation of network components, creating a cortex in dynamic steady state. Strikingly, actin turnover rates were observed to be much slower here than typical rates in cells, with actin turning over within tens of minutes rather than tens of seconds, respectively \cite{Fritzsche2013, Robin2014}. This discrepancy is likely due to the absence of dedicated actin severing proteins in the minimal system. More rapid turnover has been observed \textit{in vitro} in volume-spanning entangled actin networks where filaments were severed by cofilin and polymerization was driven by formin \cite{McCall2019}. Combining more rapid turnover with motor activity \textit{in vitro} may open a rich field of network behaviours, with complex implications for both the regulation of cortical thickness and of stress propagation and relaxation \cite{Arzash2019}.  

To build and control a system that allows actin to turn over, we can turn to the growing body of work studying the functions of various actin regulators on the single molecule or filament level. Research into the two key nucleators of cortical actin, Arp2/3 and formins \cite{Bovellan2014, Fritzsche2016, Cao2020} has uncovered new complexities in recent years. Both the processivity and the actin filament elongation rate of different formins have been shown to be regulated by the physicochemical environment, the presence of profilin, and mechanical stress \cite{Zimmermann2017, Cao2018, Suarez2015}. Even more complex co-regulation of formin with other barbed-end binding proteins is emerging \cite{Shekhar2015}. Regulation of Arp2/3 by profilin \cite{Mullins1998, Blanchoin2000} as well as by actin filament curvature \cite{Risca2012} has been known for a number of years, but the true diversity and complexity of the various isoforms of Arp2/3 is only just emerging \cite{Abella2016}. In addition, there are promoting factors that control cortex architecture by controlling both formin and Arp2/3 activity \cite{Cao2020}. Besides formin and Arp2/3, other actin nucleators such as the recently identified spire \cite{Quinlan2005} have barely been used in reconstitution experiments and may offer yet other routes towards reconstituting a minimal dynamic cortex. Actin depolymerization can equally be controlled by various factors. Disassembly of filamentous actin \textit{in vitro} is usually mediated by proteins of the ADF/cofilin family \cite{Chan2009b}. The activity of ADF/cofilin proteins has been shown to depend on cooperation with other proteins \cite{Kotila2019, Jansen2014} and on actin crosslinking \cite{Wioland2019}. ADF/cofilin also facilitates debranching in actin networks nucleated by Arp2/3 \cite{Chan2009b}, which is furthermore sensitive to force and actin filament age \cite{Pandit2020}.

\subsubsection{Reconstitution of actin cortices inside GUVs}
Controlled actin encapsulation in GUVs has proven to be a challenge. Over the years, many different methods have been explored for protein encapsulation, either based on lipid swelling \cite{Tsai2015} or on emulsion transfer \cite{Pontani2009a, Litschel2021a, Cauter2021, Bashirzadeh2021a} (reviewed in \cite{Mulla2018}). Of these, methods based on emulsion transfer are currently most successful, although the encapsulation efficiency and the ability to upscale the number of encapsulates remain to be characterized \cite{Cauter2021}. Most prior GUV studies focused on the effect of crosslink proteins and myosin motors on bulk-nucleated actin. By contrast, membrane-nucleated actin networks with turnover in GUVs remain poorly explored. Early works from the Sykes lab \cite{Pontani2009a, Murrell2011} demonstrated that Arp2/3 nucleated cortices can be reconstituted at the inner leaflet of GUVs (\cref{fig:review-natural}, green), and that such cortex-bearing vesicles reproduce aspects of the mechanics of living cells. More recently, Dürre et al. demonstrated that Arp2/3 nucleated cortices can induce local deformations of the GUV membrane by either polymerization forces alone or in combination with contractility induced by non-muscle myosin-II \cite{Durre2018}. New work from the Liu lab shows that membrane-bound Arp2/3 in combination fascin and $\alpha$-actinin is sufficient to yield ring-like membrane-bound actin networks \cite{Bashirzadeh2022}. Myosin-initiated contraction of these networks resulted in membrane constriction, thus getting one step closer to cell division. 

More extensive work, especially with myosin-driven cortices, has been performed with stable actin filaments anchored to the membrane by streptavidin or actin-binding membrane proteins. In such systems, cortical tension was shown to depend on the ratio of active versus passive crosslinkers \cite{Carvalho2013} and excessive cortical tension was shown to cause full or partial detachment of the cortex from the membrane \cite{Carvalho2013, Loiseau2016}. Recently, Litschel et al. demonstrated the formation of actomyosin rings in GUVs \cite{Litschel2021a}. However, these structures were unable to deform the GUV membrane on large length scales because they slipped on the membrane. Based on our understanding of cell division, this is likely due to (at least) three missing factors: cortex turnover, symmetry breaking between the poles and equator of the synthetic cell, and a severely limited supply of extra membrane area. Symmetry breaking is likely necessary for productive and sustained membrane deformations. There are several artificial means by which symmetry breaking could be triggered in synthetic cells. Myosin activity could, for instance, be locally light-activated by targeting either the light-sensitive myosin inhibitor blebbistatin \cite{Sakamoto2005, Linsmeier2016, Schuppler2016} (see (\cref{fig:review-natural}, orange) or myosin-II directly \cite{Yamamoto2021}. Similar approaches could be used to locally modulate the crosslink density of the actin cortex or the interaction strength of the cortex with the synthetic cell membrane. Finally, it would likely help to make GUVs shape-anisometric, for instance by using microfluidic channels \cite{Fanalista2019}.

Conceptually, building a dynamic actin cortex and pushing it towards self-assembly of a cytokinetic furrow is very attractive. Such a system would mimic many core attributes of the cortex of living animal cells. Further, the continuous nature of such a cortex would allow it take on a dual function, both as a mechanoprotective module for the synthetic cell and as a division apparatus, which sets it apart from other cytoskeletal systems such as FtsZ \cite{Barrows2021}. A life-like actin cortex offers the opportunity to test existing theoretical models of cell division and to tease out the essential functions needed for cytokinesis in living cells. On the other hand, a dynamic actin cortex will necessarily comprise more proteins and hence a higher level of complexity than one composed of stable actin filaments.  From an experimental perspective, reconstituting sustained actin turnover in combination with motor activity will in particular be challenging, as it requires fine control over both stoichiometry and activity of cytoskeletal components. 

\subsection{Engineering route: building an isolated contractile ring}
\label{section:rev-eng}

A more engineering-type approach to synthetic cell division may also be interesting: instead of building a cortex that self-organizes into a ring, one could build an isolated ring directly (\cref{fig:review-routes}, bottom). This would inherently fulfil the requirement for different activities in polar and equatorial contractility, as by definition the poles are not contractile in such a case. If a sufficient supply of long actin filaments throughout furrow ingression can be ensured, the need for controlled turnover may be diminished and the complexities of such regulated filament assembly and disassembly may be avoidable. This approach will need to address three key challenges: 1) to build an actin ring, 2) to make it contractile, and 3) to control its position and also place it mid-center such that membrane invagination rather than ring slippage occurs.

\subsubsection{How to build an isolated ring?}
\begin{figure}[h!]
	\centering
	\includegraphics[width=0.8\textwidth]{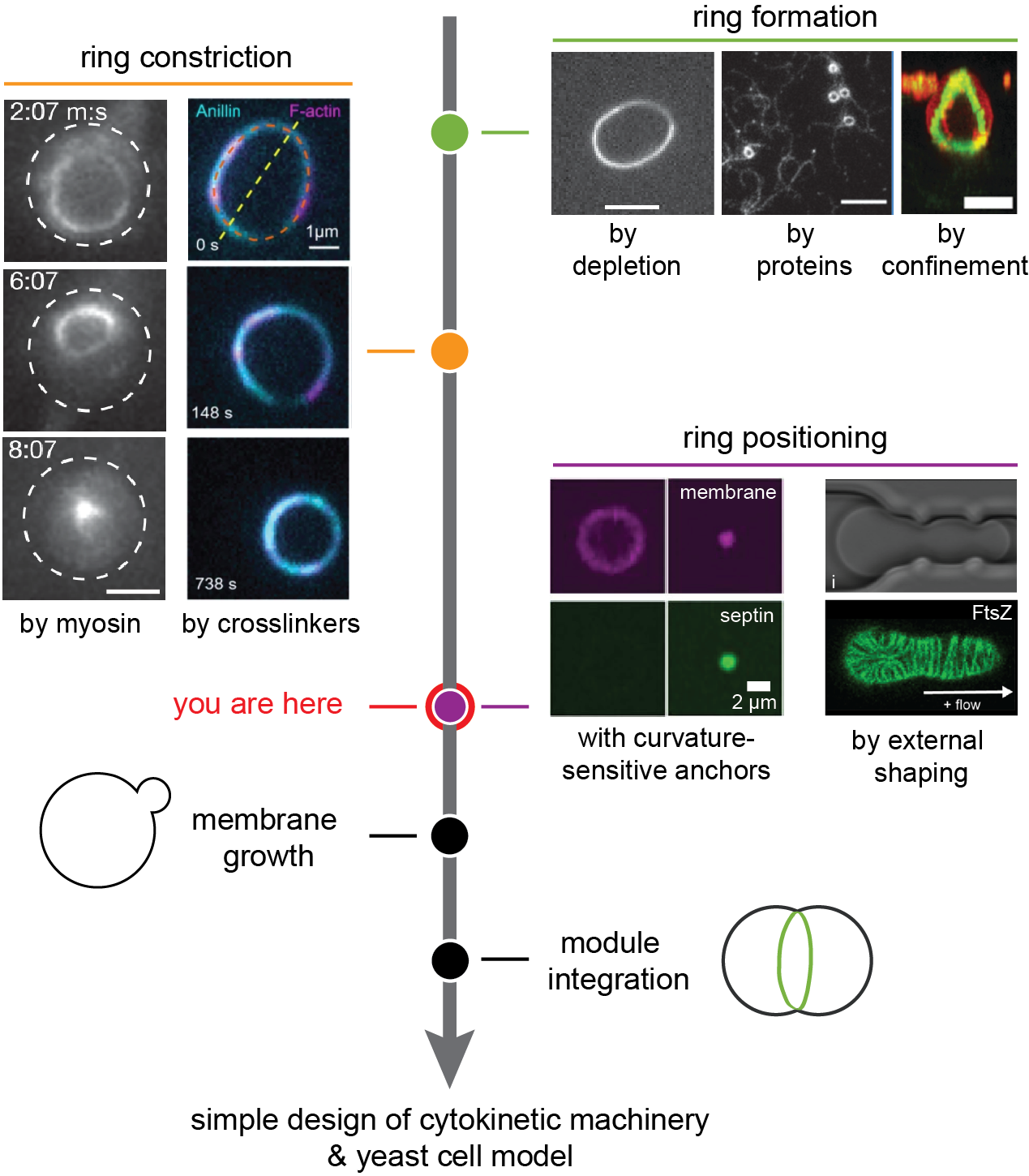}
	\caption{\textbf{Roadmap for division with contractile actin ring.} Roadmap towards synthetic cell division using a contractile actin ring. Green: actin rings can be formed by depletion interactions using macromolecular crowders \cite{Lau2009a}, by proteins that combine actin binding with curvature generation such as curly \cite{Palani2021} or simply by confinement of actin bundles \cite{Tsai2015}. Orange: constriction of actin rings can be executed using myosin motors \cite{Miyazaki2015} or using actin crosslinkers like anillin \cite{Kucera2021a}. Purple: the actin ring can be positioned using curvature-sensing anchors (left: septin binds preferentially to membranes of higher curvatures as shown with membrane-coated beads \cite{Bridges2016}) or by mechanical deformation (right: microfluidic traps deform GUVs leading to rearrangement of FtsZ rings \cite{Ganzinger2020}). In the next steps towards achieving synthetic cell division, membrane growth needs to be reconstituted, and all separate modules have to be integrated.}
	\label{fig:review-ring}
\end{figure}

Actin filaments can be bundled and bent into ringlike structures in various ways (\cref{fig:review-ring}, green). Most simply, ring formation can be induced by entropic effects through macromolecular crowding \cite{Lau2009a} or by cross-linking with multivalent ions \cite{Tang2001}. Alternatively, proteins can be used to bend actin into rings. Septins spontaneously bend actin into ringlike structures \cite{Mavrakis2014} and are recruited to the cytokinetic ring, where they cooperate with anillin in actin-membrane binding \cite{Kinoshita2002, Field2005, Piekny2008, Piekny2010, Garno2021}. Anillin itself also promotes the formation of actin rings \cite{Kucera2021a}. Further, the IQGAP fragment ‘curly’ has recently been shown to bend actin into rings on model membranes \cite{Palani2021}. The fact that all three of these proteins are enriched in the cytokinetic furrow \cite{Biro2013} suggests that these ring-forming capabilities may provide a cellular mechanism to promote successful cytokinesis.

Confinement of actin filaments inside spherical droplets or vesicles tends to promote the formation of actin rings because the confinement forces the semiflexible filaments to minimize the filament bending energy \cite{Morrison2009}. Entangled or crosslinked actin networks inside emulsion droplets and inside lipid vesicles form peripheral cortex-like networks \cite{SoaresESilva2011a, Claessens2006, Limozin2002, Bashirzadeh2021, Bashirzadeh2022, Schafer2015a}, while bundled actin forms one or more closed rings \cite{Tsai2015, Miyazaki2015, Limozin2002, Bashirzadeh2021, Bashirzadeh2022}. Single rings form when the container size is smaller than the persistence length of the actin filament or bundle \cite{Bashirzadeh2021, Bashirzadeh2022}. Recent theoretical \cite{AdeliKoudehi2019} and experimental work \cite{Litschel2021a} has shown that ring formation can be further enhanced by introducing actin-membrane adhesion. It should be noted, however, that ring formation requires a subtle balance of filament-filament and filament-wall adhesion, as well as size and stiffness of the confinement, and is not trivial to precisely control experimentally.

\subsubsection{How to contract an isolated ring?}
Contracting a once-formed actin ring can again proceed in different ways (\cref{fig:review-ring}, orange). The classical purse-string model posits a well-organized cytokinetic ring that closes by myosin-mediated translocation of actin filaments \cite{Schroeder1968, Schroeder1972}. Although this model does not appear to hold in all cell types \cite{Ma2012, Reichl2008, Fishkind1993}, recent superresolution and electron microscopy showed convincing evidence that it does apply at least in some cell types \cite{Fenix2016, Henson2017}. Contracting actin-myosin rings have been succesfully reconstituted on supported lipid bilayers \cite{Palani2021} and inside water-in-oil droplets \cite{Miyazaki2015} and GUVs \cite{Litschel2021a}. The efficiency of ring closure is likely determined by the orientation and arrangement of the actin filaments in the ring, which can be tuned by crosslinker composition and concentration \cite{Bashirzadeh2021, Lenz2014, Lenz2012, Ennomani2016, Schwayer2016}.

Alternatively, ring contraction may be driven by mechanisms that do not require molecular motors. For instance, anillin was recently shown to drive actin bundle contraction even though it is a passive crosslinker \cite{Kucera2021a}. Contraction was attributed to an energetically driven process whereby actin filaments increase their overlap as long as energy can be gained by accumulating diffusive crosslinkers in the overlap region \cite{Kucera2021a}. This mechanism was enhanced when anillin was combined with actin depolymerization. Since contraction driven by passive crosslinkers does not consume energy from an external energy source such as ATP, it can only bring the system into a configuration of minimal free energy, at which point rearrangement will stop \cite{Odde2015}. Intriguingly, recent theoretical modelling \cite{Chen2020} suggests that a crosslinker that consumes ATP to unbind from actin filaments, but does not actively translocate them like myosin, could in principle induce contraction indefinitely. In this case, the consumption of an energy carrier breaks detailed balance in the system, and in combination with the asymmetric mechanical properties of actin, overall contractile forces can arise.

\subsubsection{How to keep an isolated ring in place?}
Although contractile actin rings have been successfully reconstituted inside GUVs, so far none of these efforts have yielded anything close to furrow-like membrane invaginations. The rings either detached or slipped along the membrane upon myosin activation \cite{Carvalho2013, Miyazaki2015, Litschel2021a, Loiseau2016}, at best producing rare instances of slight membrane deformation \cite{Litschel2021a}. In cells, positioning of the cytokinetic ring is ensured by a complex and poorly understood interplay between the actin and microtubule cytoskeleton, local changes in lipid composition, and soluble signalling molecules \cite{Rizzelli2020, Lancaster2014}. Reconstituting this interplay in GUVs seems too technically challenging to be expected in the coming years. We therefore expect that simpler, if less biological, solutions may be more promising. To the best of our knowledge, no such efforts have been reported so far. However a few options present themselves (\cref{fig:review-ring}, purple): curvature-sensing or –inducing scaffolding proteins such as septins \cite{Bridges2016} or I-BAR-domain proteins \cite{Tsai2018, Quinones2010} may help in templating a furrow and inhibiting slippage of contractile actin rings. These proteins may have to be combined with more engineering-type solutions designed to deform GUVs from the outside, either by confinement in traps \cite{Fanalista2019, Ganzinger2020} or by membrane-binding complexes \cite{Bae2019b, Czogalla2015, Steinkuhler2020}.

Building an isolated contractile actin ring in principle offers an elegant way to drive synthetic cytokinesis. Formation of such a ring requires only few components and tuning ring contractility is certainly subtle, but most likely achievable. The biggest technical challenge in this approach is to localize the ring at the equator and keep it in place during contraction so as to foster productive membrane deformation. On a more conceptual level, reconstituting isolated contractile rings likely will not bring us much insight into the mechanisms of cytokinesis in animal cells. It may however be a valid strategy for understanding mechanisms in yeast cytokinesis, in tandem with top-down work on yeast cell ghosts \cite{Mishra2013a}. 

\section{Involving the membrane}
So far, we have largely ignored an important assumption in the key requirements that we set out earlier, which is that the GUV membrane and actin cortex are intrinsically coupled. However, it is far from trivial that actomyosin contraction is followed by deformation of the cellular membrane. While actomyosin networks and membranes have separately been thoroughly investigated by biophysicists, their interplay has received much less attention and presents a crucial challenge to address in the coming years. 

\begin{figure}[h!]
	\centering
	\includegraphics[width=0.8\textwidth]{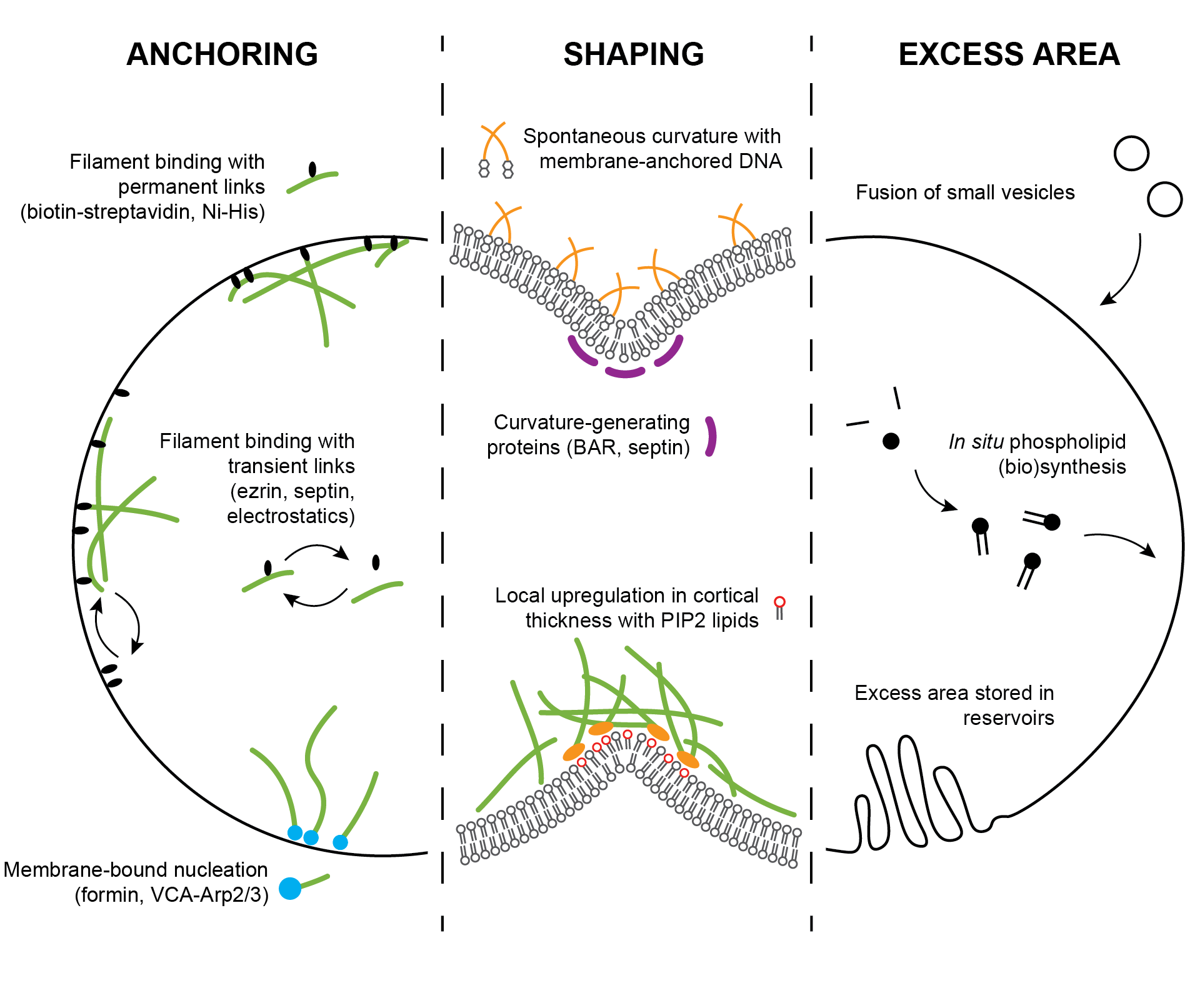}
	\caption{\textbf{Membrane engineering for synthetic cell division.} Schematic overview of possibilities for membrane design. Anchoring of the actin cortex (left) can be done either via filament nucleation from the membrane, or via filament binding to the membrane. Binding can be done using strong permanent linkers, or using weaker transient links. Membrane shaping (middle) can be done by generating spontaneous curvature, for example with membrane-bound DNA nanostars \cite{Franceschi2021} or by physiological curvature-generating proteins such as BAR proteins \cite{McMahon2015} or septin \cite{Beber2019b}. Otherwise, when lipids can be spatially separated, local elevation of PIP$_2$ levels can increase cortical thickness via regulating actin nucleation and severing proteins. To provide excess area during cytokinesis (right), new membrane area can be added by fusion of small vesicles or by \textit{in situ} synthesis of phospholipids. Alternatively, membrane area could be stored in reservoirs that become accessible upon furrow ingression.}
	\label{fig:review-membrane}
\end{figure}

\subsection{Membrane-cortex anchoring}

\textit{In vivo}, a multitude of cytoplasmic proteins is known to be involved in actin-membrane adhesion, many of which have binding sites for both actin and membrane lipids. These proteins include ERM (ezrin, radixin, moesin)-proteins, myosin1b, anillin and septins \cite{Taubenberger2020, Prosperi2015, Russo2021, Michie2019}. How these proteins cooperate in adhesion and how they are spatially organized at the membrane remains elusive. Electron microscopy and superresolution microscopy have revealed that the distance between the filamentous actin and the plasma membrane is surprisingly large, ranging from 10 to 20 nm in the cell cortex of animal cells \cite{Clausen2017} and from 60 to 160 nm in the cytokinetic ring of fission yeast \cite{Swulius2018, McDonald2017}. It is unclear how this large gap, which is often wider than the distance that known linker proteins span, arises. There is evidence that the actin cortex itself is stratified, with myosin filaments being restricted towards the cytoplasmic side of the cortex due to steric exclusion from the dense cortex \cite{TruongQuang2021}. Interestingly, a recent \textit{in vitro} reconstitution study showed that actin-myosin networks on supported lipid bilayers spontaneously self-organize into radial actin structures (asters) with myosin at the core and layered atop to relieve steric constraints \cite{Das2020}.   

Mechanical measurements on cells indicate that the cortex adheres to the membrane via a high density of weak links. With optical tweezers, one can pull membrane tubes from cells with membrane-bound beads. These tubes can easily be moved over the cell surface \cite{Sheetz2001}, indicating that the membrane easily zips off the cortex and quickly rebinds. Various tube pulling experiments have shown that the force required for tube extrusion is dependent on the levels of ezrin \cite{Paraschiv2021} and phosphatidylinositol-4,5-bisphosphate (PIP$_2$) lipids \cite{Raucher2000}. PIP$_2$ lipids specifically interact with many actin-binding proteins including ezrin (reviewed in \cite{Janmey1999}). In \textit{S. Pombe} cells, PIP$_2$ depletion causes sliding of the cytokinetic ring, indicating that PIP$_2$-dependent actin-membrane adhesion is essential for anchoring of the ring \cite{Snider2016}. Although PIP$_2$-protein interactions are individually weak, their high density collectively causes a tight yet dynamic seam between bilayer and cytoskeleton. 

In stark contrast to the reversible actin-membrane binding observed \textit{in vivo}, \textit{in vitro} reconstitution efforts have mostly relied on anchoring interactions with unphysiologically high binding affinity (\cref{fig:review-membrane}, left). Many studies used either direct coupling of biotinylated actin filaments to biotinylated lipids via streptavidin \cite{Carvalho2013, Simon2018, Litschel2021a} or indirect coupling using His-tagged actin-binding proteins coupled to Ni-NTA lipids \cite{Shrivastava2015, Loiseau2016}. These bonds are virtually permanent and unbreakable \cite{Rico2019, Nye2008, Raghunath2019}. As described above, actin-myosin cortices anchored in this manner typically detach from the membrane upon myosin activation \cite{Loiseau2016,Carvalho2013}. In two studies with high anchor density, the acto-myosin cortex did remain attached to the membrane upon contraction, but it slid towards one side so the membrane was only minimally deformed \cite{Carvalho2013,Litschel2021a}. Cortex slippage is likely due to the fluid nature of the lipid bilayer membrane. Actin and microtubule gliding assays with motors proteins anchored onto supported lipid bilayers have shown that motor activity is accompanied by lipid slippage  \cite{Grover2016, Pernier2019}. The interplay between the dynamics of the actin cortex and the dynamics of the lipids is complicated. Adhesion to the actin cortex slows down lipid diffusion \cite{Schneider2017,Heinemann2013}, while myosin-driven actin cortex contraction can actively cluster lipids into microdomains \cite{Vogel2017, Koster2016, Honigmann2014,Rao2014, Gowrishankar2012, Liu2006}. Altogether, it remains poorly understood what conditions are necessary for the actin cortex to remain stably anchored and cause sustained membrane deformation.  

Dynamic actin-membrane linkage has so far been reconstituted only on supported lipid bilayers. Using ezrin recruited to the bilayer via PIP$_2$ lipids, indeed a dynamic actin network was created that could be remodelled by passive filament cross-linkers \cite{Schon2019}. Bead tracking microrheology showed that ezrin serves as a dynamic cross-linker for the membrane-attached actin layer with the network stiffness being controlled by the pinning point density \cite{Noding2018}. Ezrin-anchored actin filaments could diffuse over the membrane but longer filaments were immobilized, being pinned by a larger number of actin-membrane links \cite{Mosby2020}. This indicates that collective binding with transient links can fix cytoskeletal structures in place on top of a fluid membrane. Other promising candidates for \textit{in vitro} transient actin binding are septins and anillin. Septins themselves can bind to membranes and self-assemble into filamentous scaffolds \cite{Szuba2021}. Membrane binding is curvature-sensitive \cite{Beber2019b, Cannon2019}, which renders septins interesting candidates for spatially controlling actin organization in synthetic cells. In solution, septins can bind and crosslink actin filaments into curved bundles \cite{Mavrakis2014}. This could explain the role of septins in the formation and stabilization of contractile acto-myosin rings observed \textit{in vivo} \cite{Mavrakis2014}. However, the simultaneous interplay of septins with lipid membranes and actin has yet to be reconstituted \textit{in vitro}. Like septins, also anillin possesses both actin-binding and membrane-binding domains. Anillin has been shown by reconstitution to be able to anchor actin filaments to lipid membranes in a RhoA-dependent manner \cite{Sun2015}. In combination with anillin’s ability to bundle and constrict actin rings via condensation forces \cite{Kucera2021a}, it would be interesting to explore anillin's ability to promote synthetic cell division. Besides protein-based binding, actin filaments can also be bound to lipid membranes by electrostatic interactions that can be tuned by the choice of ions, offering an alternative route for studying and modulating transient actin-membrane binding \cite{Schroer2020}.

Besides actin-membrane linkers, also membrane-localized actin nucleation contributes to cortex-membrane adhesion. The main nucleators of cortical actin filaments \textit{in vivo} are Arp2/3 and formin \cite{Bovellan2014}. Arp2/3 in combination with membrane-bound nucleation promoting factors such as WASP are responsible for the formation of branched actin filament arrays, whereas formins nucleate linear filaments. Actin nucleation has been successfully reconstituted \textit{in vitro} both with formins, often for simplicity with constitutively active mutants \cite{Zigmond2004}, and with Arp2/3, often activated by WASP fragments such as VCA \cite{Romero2004, Pontani2009a, Guevorkian2015}. Actin turnover can be introduced by addition of severing proteins such as ADF/cofilin \cite{Bleicher2020}. 

It is unknown how filament nucleation in conjunction with actin-membrane anchoring by dynamic linker proteins such as ezrin will influence the ensemble mechanics of the actin-membrane composite. Tailoring actin-based division machineries towards synthetic cell division will require careful tuning of the cortex itself, the anchoring strategy, and also the membrane physico-chemical properties.

\subsection{Membrane engineering}
The membrane should not be considered just a passive player in cytokinesis. In contrast, membrane properties can be exploited to aid cytokinesis, for example by shaping the contractile network (\cref{fig:review-membrane}, middle). \textit{In vivo}, the plasma membrane in the cleavage furrow has a distinct lipid composition that is thought to contribute to cytokinesis by biochemical signalling and perhaps also by induction of spontaneous curvature \cite{Ng2005}. Elevated PIP$_2$ levels at the cleavage furrow probably contribute to furrow ingression by recruiting anillin, septins and ERM-proteins \cite{Cauvin2015}. Furthermore, PIP$_2$-mediated signalling promotes the formation and maintenance of a stable actin cortex by promoting actin nucleation and slowing down actin filament severing via actin regulatory proteins \cite{Logan2006}. Other membrane compontents such as gangliosides and cholesterol also accumulate in the cleavage furrow, where they regulate and bind the cortex \cite{Cauvin2015}. In addition, the distribution of phosphatidylethanolamine (PE) lipids over the two bilayer leaflets changes significantly during cell division: while PE lipids reside in the inner leaflet during interphase, they are exposed in the outer leaflet of the cleavage furrow during cytokinesis \cite{Emoto2001}. This asymmetric distribution of PE lipids has been shown to be important for disassembly of the contractile ring after cytokinesis \cite{Emoto2001}. It is possible that the specialized lipid composition of the cleavage furrow also directly affects cytokinesis by changing the mechanical properties of the membrane, but this remains to be shown.

For engineering artificial cell division, it could be useful to exploit known mechanical effects of lipids. An important characteristic of lipid bilayers is that asymmetries between the two membrane leaflets give rise to membrane spontaneous curvature. Asymmetries can be generated in many different ways (reviewed in \cite{Bassereau2018}), such as by different lipid compositions or different numbers of lipids in the two leaflets \cite{Dasgupta2018}, proteins binding to one leaflet\cite{Steinkuhler2020}, membrane-anchored DNA oligos inserting into one leaflet\cite{Franceschi2021}, or different solutes on both sides of the membrane \cite{Karimi2018}. In the context of actomyosin-based synthetic cell division, spontaneous curvature effects could be exploited for spatial control and symmetry breaking. Binding of proteins to the outer leaflet of vesicles can be used to make vesicles dumbbell-shaped and to constrict and even split the neck \cite{Steinkuhler2020}. Generation of negative membrane curvature could be used to locally recruit septins, which selectively bind to membrane areas with micrometric curvature \cite{Bridges2016, Beber2019b}. In addition, membrane-binding proteins that not only sense, but also generate curvature could be used, such as BAR-domain proteins. I-BAR proteins were shown to directly bind to actin in fission yeast \cite{Quinones2010} and are therefore interesting candidates for promoting actomyosin-driven membrane invagination. Interestingly, I-BAR domain proteins promote ezrin enrichment in negatively curved membrane protrusions \cite{Tsai2018}, providing further prospects for boosting membrane invagination \textit{in vitro}. 

\subsection{Addition of new membrane area}
To create two daughter cells from a single mother cell, assuming spherical geometry, the cell surface area has to increase by ~28\% \cite{Frey2021}. \textit{In vivo}, this extra membrane area is delivered to the cleavage furrow by targeted endosomal transport \cite{Shuster2002}. This mechanism does not only lead to a local area increase, but also allows fast and localized delivery of specific lipids and regulatory proteins (reviewed in \cite{Schiel2013}). For reconstitution of cell division, various strategies can be followed to increase the membrane area (\cref{fig:review-membrane}, right). First, GUV membranes can be grown by external addition of small unilamellar vesicles (SUVs) that can be forced to fuse with the GUV using fusogenic peptides, DNA, or charge-based interactions \cite{Marsden2009, Lira2019, Deshpande2019, Dreher2021}. Second, lipid membranes can be grown by \textit{in situ} synthesis of lipids from their precursors. Examples are non-enzymatic reactions from synthetic reactive precursors \cite{Kurihara2011} or enzyme-catalysed biosynthesis using either purified proteins \cite{Exterkate2018} or \textit{in vitro} transcription-translation \cite{Blanken2020}.
Although there is evidence that mammalian cells do not use area reservoirs, such as microvilli, to supply extra membrane area for division \cite{Schroeder1981, Wong1997}, this mechanism could be exploited for engineering division in a synthetic cells. Asymmetries between the two leaflets of the bilayer generated by different means (see preceding section) can be used to store excess area in membrane tubes and buds \cite{Steinkuhler2020, Karimi2018, Bhatia2018, Bhatia2020}. Low forces suffice to access these reservoirs \cite{Karimi2018, Bhatia2020}.
To achieve synthetic cell division, it will be important to match timing of membrane areal growth with the timing of actin-driven constriction. To achieve multiple cycles of division, it will moreover be important to build in a mechanism to maintain lipid homeostasis.

\section{Challenges ahead}
In the past decades, our knowledge of cell division and its molecular actors has increased tremendously. To understand the physical mechanisms governing actomyosin-driven cell division, focus is put increasingly on bottom-up reconstitution experiments. Bulk and SLB experiments have helped us to understand mechanics of active actomyosin networks in 3D and 2D. However, translating these insights to the process of cell division is not trivial. To summarise, we list here the critical challenges that need to be overcome before we can reconstitute a minimal version of actin-driven cell division. 

First, we need to understand how actin network contraction is sustained to drive division all the way. This will require myosin activity working in concert with actin turnover. While activity and turnover have been studied to great extent individually, we still have minimal understanding how they together govern actin network mechanics and contractility. Not only is this a challenging system to understand from a physical and biological perspective, also from an experimental perspective it is difficult to recapitulate, as it involves a large number of components whose concentration and activity need to be tightly controlled. More \textit{in vitro} work in this direction, both in 3D and 2D, will be essential to explore the parameter space. 

Second, it remains elusive how the actomyosin network should be anchored onto the membrane in order to achieve membrane deformations. A multitude of anchoring strategies has been developed and investigated, but only minimally in combination with a deformable membrane. Combined with our limited understanding of cortex-membrane molecular organization \textit{in vivo}, this might prove one of the most important challenges. Future studies need to focus on understanding the influence of linker density and strength, as well as membrane composition and organization. In addition, the respective roles of filament-membrane linkers and membrane-bound nucleators need to be investigated.

Third, attention must be paid to the supplying of extra membrane area during constriction. Additional area can be present in membrane reservoirs, be synthesized, or be added by fusion of small vesicles. However, none of these approaches have to our knowledge been co-reconstituted with actin-driven contraction and resulting membrane deformation. 

Fourth, there is to date only a minimal body of work on contractile actomyosin networks in GUVs. Confining the system in GUVs requires that all components are encapsulated in the right concentration and stoichiometric ratio, while preserving functionality. Although there are numerous GUV formation techniques, they have been minimally characterized for their potential to encapsulate complex mixtures of biochemically active components. More work in this direction is crucial to perform controlled reconstitution in GUVs, but also to be able to extrapolate findings from bulk and SLB experiments to vesicle systems.

Fifth, spatial and temporal control of the components and their activity is crucial. On the short term, some of the involved challenges may be by-passed by taking a semi-autonomous approach to synthetic cell division. For example, optogenetics, external mechanical or chemical cues, or fusion-based delivery of components with small vesicles provide handles to control the system even after encapsulation of the components inside GUVs. However, if the goal is to create a synthetic cell that divides fully autonomously, reconstitution will be more complicated, requiring for example feedback loops, signalling molecules and internal clocks. 

As a concluding remark, we note that the most pressing challenges to achieve \textit{in vitro} actin-driven cell division require integration of modules. Only when actomyosin studies meet membrane biophysics, when myosin motor activity is combined with actin turnover, and when protein biochemistry becomes integrated in GUV formation, we can start thinking about reconstituting cell division. In the coming years, perspectives from experimental work, theoretical studies and simulations need to be combined to guide future work with the ultimate goal to develop a full understanding of actin-driven synthetic cell division.

\section{Acknowledgements}
We would like to thank Ilina Bareja, Gerard Castro-Linares and Fred MacKintosh for useful discussions about actin cross-linkers. We acknowledge financial support from The Netherlands Organization of Scientific Research (NWO/OCW) Gravitation program Building A Synthetic Cell (BaSyC) (024.003.019). 


\bibliography{review}

\end{document}